\documentclass[trackchanges, twocolumn]{aastex701}

\usepackage{graphicx} % Required for inserting images
\usepackage{xcolor} % For \textcolor command
\usepackage{amssymb}
\usepackage{amsmath}
\usepackage{physics}
\usepackage{comment}
\usepackage{gensymb}
\usepackage{placeins}
\usepackage{amsfonts}
\usepackage{hyperref}

\usepackage{enumerate}

\begin{document}

%%%%%%%%%%%%%%%%%%%%%%%%%%%%%%%%%%%%%%%%%%%%%%%%%%%%%%%%%%%%%%%%%%%%%%%
\title{Modeling the Milky Way Circumnuclear Disk: Rosettes and Rings}

\author[orcid=0009-0000-7650-7164,sname='Ukani']{Asad Ukani}
\email{asadukani2028@u.northwestern.edu}
\affiliation{Center for Interdisciplinary Exploration and Research in Astrophysics (CIERA), Northwestern University, Evanston, IL 60208, USA}
\affiliation{Department of Physics and Astronomy, Northwestern University, Evanston, IL 60208, USA}
\email[show]{asadukani2028@u.northwestern.edu}  

\author[orcid=0000-0001-8986-5403, sname='Elena Murchikova']{Lena Murchikova}
\affiliation{Center for Interdisciplinary Exploration and Research in Astrophysics (CIERA), Northwestern University, Evanston, IL 60208, USA}
\affiliation{Department of Physics and Astronomy, Northwestern University, Evanston, IL 60208, USA}
\affiliation{School of Natural Sciences, Institute for Advanced Study, 1 Einstein Drive, Princeton, NJ 08540, USA}
\email{lena@northwestern.edu}

\author[orcid=0000-0001-9300-354X,sname='Mark Gorski']{Mark D. Gorski}
\affiliation{Center for Interdisciplinary Exploration and Research in Astrophysics (CIERA), Northwestern University, Evanston, IL 60208, USA}
%\affiliation{Department of Physics and Astronomy, Northwestern University, Evanston, IL 60208, USA}
\email{mark.gorski@northwestern.edu}  


%%%%%%%%%%%%%%%%%%%%%%%%%%%%%%%%%%%%%%%%%%%%%%%%%%%%%%%%%%%%%%%%%%%%%%%


\begin{abstract}
The Milky Way Galactic Center hosts a $\sim4\times10^{6}\,M_\odot$ supermassive black hole (SMBH), Sagittarius A* (Sgr~A*). The dominant structures in its immediate vicinity are the nuclear star cluster (NSC), whose enclosed mass at 2\,pc is approximately half that of the SMBH, and the circumnuclear disk (CND)/ring, which extends between $\sim0.5\,$pc and $\sim3\,$pc from Sgr~A* and is the largest reservoir of molecular gas in this region. Existing models of the CND commonly use one circular orbit to describe the motion of its gas. Here, we explore a much broader range of models. In the combined potential of Sgr~A* and the NSC, we consider non-Keplerian rosette orbits as well as a circular disk, which is formed using a finely spaced set of concentric rings. For both systems, we test various inner/outer radii, inclinations, and position angles, sampling a total of $\sim3.3 \times 10^{5}$ models. We then conduct mock observations of all models to construct velocity maps, which we compare with HCN ($J=1{-}0$) observations of the CND. We find that the best-fitting model is a circular disk with inner and outer radii of $1.0\,$pc and $2.9\,$pc, an inclination of $i=60\degree$, and a position angle of $\text{PA} = 35\degree.$  
\end{abstract}

\keywords{\uat{Active galactic nuclei}{16} --- \uat{Astrophysical black holes}{98} --- \uat{Galactic center}{565} --- \uat{Interstellar medium}{847} --- \uat{Galaxy circumnuclear disk}{581}}


%%%%%%%%%%%%%%%%%%%%%%%%%%%%%%%%%%%%%%%%%%%%%%%%%%%%%%%%%%%%%%%%%%%%%%%
\section{Introduction}\label{sec:intro}
\begin{figure}
\centering
\includegraphics[width=0.47\textwidth]{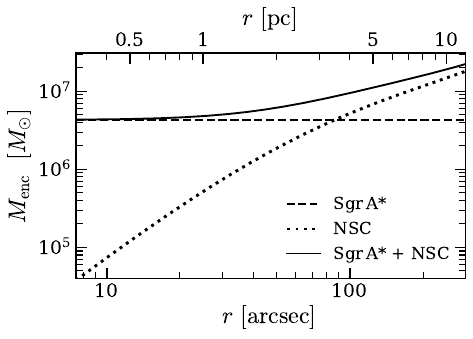} 
\caption{
Enclosed mass in the Milky Way's Galactic Center within radius $r$ from Sgr~A*. The dashed line indicates the SMBH mass ($M_{\mathrm{SgrA*}}$), which does not change with distance. The dotted line indicates the stellar mass of the nuclear star cluster ($M_{\mathrm{NSC}}(r)$), obtained from Equations \ref{eq:mass_dens_profile} and \ref{eq:M_nsc_encl} \citep{Chatzopoulos_2015}.
The solid line is the sum of these components: $M_{\mathrm{enc}}(<r)$ = $M_{\mathrm{SgrA*}} + M_{\mathrm{NSC}}(r)$. The radius $r$ is marked in both pc (top) and arcsec (bottom).
}
\label{fig:enclosed_mass}
\end{figure}

The inner few parsecs of the Milky Way Galactic Center (GC) present a complex environment. At the center, there is a $\sim4\times10^{6}\,M_\odot$ supermassive black hole (SMBH), Sagittarius A* (Sgr~A*; \cite{Gravity2019, Do2019}), which is surrounded by two dominant structures: the circumnuclear disk/ring (CND) and the nuclear star cluster (NSC). The CND is a ring of dense molecular gas and dust, which extends from $\sim0.5\,$pc to $\sim3\,$pc and has a mass of $\sim10^{4}{-}10^{6}\,M_\odot$ \citep{scoville_2005, Etxaluze2011, Oka_2011, Requena-Torres2012, Lau2013, Liu2013, Hsieh2018, Tsuboi_2018, James2021}. As the largest reservoir of molecular gas in the inner few parsecs of the Galactic Center, it is the most important source for supplying molecular gas to the central $0.5\,$pc and for regulating the multiphase accretion flow around Sgr~A* \citep{Montero-Castaño_2009, Liu_2012, Moser_2017, Hsieh_2019}. Moreover, the disk may be a potential site or source for star formation \citep{morris_1996_review, Yusef_zadeh_2008b}. Therefore, understanding the dynamics of the CND is crucial for advancing Galactic Center studies. The NSC is centered on Sgr~A* and extends approximately beyond 10\,pc in radius \citep{schodel_2014_review}. Within 2\,pc, its enclosed mass is equal to about half that of the black hole \citep{Chatzopoulos_2015, schodel_2018_II} (Figure~\ref{fig:enclosed_mass}). The gravitational potential of the NSC is thus expected to affect the dynamics of the CND. 

Previous studies find that the CND rotates about Sgr~A* at a velocity of $\sim110\,\mathrm{km\,s^{-1}}$, is inclined by $i = 50\degree{-}75\degree$ relative to the plane of the sky, its projected major axis has a position angle (PA) of $25\degree{-}30\degree$ east of north, and its plane of rotation at a projected radius of $1.7$\,pc is tilted by $5\degree$ relative to the Galactic plane \citep{Genzel_1989, Genzel_2010, Ferriere_2012}. The CND is commonly modeled by assuming that its gas is moving on one (or a few) dominant circular orbit(s) under the influence of the SMBH mass combined with that of the enclosed stellar cluster \citep{Oka_2011, SW_2014, Hsieh_2017, Goicoechea_2018, Tsuboi_2018}. The radius of such a ring is chosen to be either the distance where the disk's emission is close to its maximum value or the radius of the orbit that best fits the prominent components of the CND.

%%%%%%%%%%%%%%%%%%%%%%%%%%%%%%%%%%%%%%%%%%%%%%%%%%%%%%%%%%%%%%%%%%%%%%
%%%%%%%%%%%%%%%%%%%%%%%%%%%%%%%%%%%%%%%%%%%%%%%%%%%%%%%%%%%%%%%%%%%%%%

Here, we present the first comprehensive study of the CND that models its rotation using both a circular disk and non-Keplerian rosette orbits. The rosettes are non-circular orbits in the combined SMBH and NSC potential that exhibit apsidal precession. We test all our models against an observational velocity map of the CND \citep{Hsieh_2021}. For every model, we create a mock-observational velocity map using the same beam parameters as the observations. We then evaluate models based on their deviation from the observed velocities and their spatial coverage of the CND's on-sky area. The combined number of models we consider in this work is $\sim3.3\times10^{5}$, making it the most extensive modeling study of the CND to date. 

This paper is organized as follows. In Section~\ref{sec:models}, we present our models of the CND. Section~\ref{sec:obs} details how we compare these models to observations, explaining our method for constructing mock observations and the metric we use to evaluate model quality. In Section~\ref{sec:discussion}, we report the best-fit parameters for every model type and compare the models. We conclude in Section~\ref{sec:conclusion}. 

%%%%%%%%%%%%%%%%%%%%%%%%%%%%%%%%%%%%%%%%%%%%%%%%%%%%%%%%%%%%%%%%%%%%%%
\begin{figure}
\centering
\includegraphics[width=0.47\textwidth]{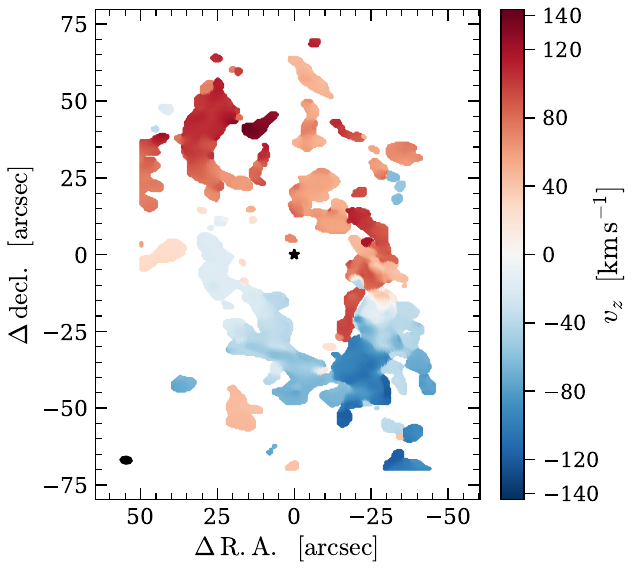} 
\caption{Emission-weighted mean velocity map (moment-1) of the CND in the HCN ($J=1{-}0$) line, constructed using observational data with a signal-to-noise ratio $\sigma \geq 20$ from \cite{Hsieh_2021}. The data extend farther than the plotted region, but we crop the map to retain only the most prominent features of the CND. The synthesized beam, which has an FWHM of $3.''69 \times \,2.''34$ at a PA of $85\degree.2$, is depicted in the bottom left corner. The black star indicates the location of Sgr~A*.}
\label{fig:obs_panel_fig}
\end{figure}
%%%%%%%%%%%%%%%%%%%%%%%%%%%%%%%%%%%%%%%%%%%%%%%%%%%%%%%%%%%%%%%%%%%%%%%

%%%%%%%%%%%%%%%%%%%%%%%%%%%%%%%%%%%%%%%%%%%%%%%%%%%%%%%%%%%%%%%%%%%%%%%
\section{Modeling}\label{sec:models}
%%%%%%%%%%%%%%%%%%%%%%%%%%%%%%%%%%%%%%%%%%%%%%%%%%%%%%%%%%%%%%%%%%%%%%%

%%%%%%%%%%%%%%%%%%%%%%%%%%%%%%%%%%%%%%%%%%%%%%%%%%%%%%%%%%%%%%%%%%%%%%%
\subsection{Gravitational Potential in the Galactic Center}\label{sec:NSC_pot}
%%%%%%%%%%%%%%%%%%%%%%%%%%%%%%%%%%%%%%%%%%%%%%%%%%%%%%%%%%%%%%%%%%%%%%%
Throughout this paper, we assume a central black hole mass of $M_{\textrm{SgrA*}} = 4.30\times10^{6}\,M_\odot$ and a distance to the Galactic Center of $R_{0} = 8.28\,$kpc \citep{GRAVITY_2022}. If we use the values from \cite{Do2019} instead, our estimations would change by $\sim5\%$. 

We adopt the NSC mass density profile ($\rho_{\text{NSC}}$) from the two-component spherical $\gamma$-model of \cite{Chatzopoulos_2015}:
\begin{equation}
\label{eq:mass_dens_profile}
        \rho_{\textrm{NSC}}(r) = \sum_{i=1}^{2} \frac{3 - \gamma_{i}}{4\pi} \cdot \frac{M_{\mathrm{NSC},i} \cdot a_{i}}{r^{\gamma_{i}} \, (r + a_{i})^{4 - \gamma_{i}}},
\end{equation}
where $r$ is the deprojected distance from Sgr~A*, and the NSC parameters are $M_\textrm{NSC,\,1} = 2.7\times10^7\,M_\odot$ \citep{schodel_2014, Fritz_2016, Feldmeier-Krause_2025}, $a_1 = 3.97\,$pc, $\gamma_1 = 0.51$, $M_\textrm{NSC,\,2} = 2.8\times10^9 \,M_\odot$, $a_2 = 95.38\,$pc, and $\gamma_2 = 0.07$.

The enclosed mass of the NSC within radius $r$ $(M_{\mathrm{NSC}}(r))$ can be obtained by integrating Equation~\ref{eq:mass_dens_profile} (Figure~\ref{fig:enclosed_mass}):
\begin{equation}\label{eq:M_nsc_encl}
M_{\mathrm{NSC}}(r) = \int^r_0 \rho_{\text{NSC}}(r') 4 \pi r'^2 \,dr'.
\end{equation}

The combined gravitational potential of the SMBH and the NSC has the form
\begin{equation}
\label{eq:tot_gravPot}
\begin{aligned}
    \Phi(r) = -\frac{GM_{\textrm{SgrA*}}}{r} &- \sum_{i=1}^{2} \frac{GM_{\textrm{NSC},i}}{a_i \cdot (2 - \gamma_i)} \\
    &\times \left[ 1 - \left( \frac{r}{r + a_i} \right)^{2 - \gamma_i} \right].
\end{aligned}
\end{equation}

%%%%%%%%%%%%%%%%%%%%%%%%%%%%%%%%%%%%%%%%%%%%%%%%%%%%%%%%%%%%%%%%%%%%%%%
\subsection{Circular Disk Model}\label{sec:kep_methods}
%%%%%%%%%%%%%%%%%%%%%%%%%%%%%%%%%%%%%%%%%%%%%%%%%%%%%%%%%%%%%%%%%%%%%%%
The Circular Disk model describes the CND as a smooth disk that is formed using a set of concentric rings, which are separated by a width of $\Delta\,r\sim10^{-3}\,$pc. The disk's deprojected inner radius is labeled $r_{\textrm{in}}$ and its outer radius $r_{\textrm{out}}$. In its orbital plane, each individual ring rotates about the dynamical center in the counterclockwise direction with a rotation velocity 
\begin{equation}
v(r) = \sqrt{\frac{G M_\mathrm{enc}(<r)}{r}},
\end{equation}
where $M_{\mathrm{enc}}(<r) = M_{\mathrm{SgrA*}} + M_{\mathrm{NSC}}(r),$ and $M_{\mathrm{NSC}}(r)$ is calculated using Equation~\ref{eq:M_nsc_encl}. 

The range of model parameters we explore is listed in Table~\ref{tab:params_sampled}. Orientation angles are discussed in Section \ref{sec:mock_obs}.

%%%%%%%%%%%%%%%%%%%%%%%%%%%%%%%%%%%%%%%%%%%%%%%%%%%%%%%%%%%%%%%%%%%%%%%
\subsection{Rosette Models}\label{sec:ros_methods}
%%%%%%%%%%%%%%%%%%%%%%%%%%%%%%%%%%%%%%%%%%%%%%%%%%%%%%%%%%%%%%%%%%%%%%%
In the central potential $\Phi(r)$ from Equation~\ref{eq:tot_gravPot}, any bound non-circular orbit will exhibit apsidal precession and trace a rosette-like trajectory \citep{BT_2008}. As the potential $\Phi(r)$ is spherically symmetric, the orbit remains confined to a plane. There is no exact solution for a rosette orbit in this potential, so we solve for it numerically using the $4^{\text{th}}$-order Runge${-}$Kutta method. We initialize a counterclockwise orbit at pericenter $r_{\mathrm{in}}$ with angular momentum $\ell$ and velocity $v_{\text{init}}$
\begin{equation}
\ell^{2} = -2 \frac{\Phi(r_{\mathrm{in}}) - \Phi(r_{\mathrm{out}})}{r_{\mathrm{in}}^{-2}\, - \,r_{\mathrm{out}}^{-2}}, \qquad
v_{\text{init}} = \frac{\ell}{r_{\mathrm{in}}} \, ,
\end{equation}
where $r_{\text{out}}$ is the apocenter of the orbit.

We consider two types of rosette models:

\textit{The Million-Years-Rosette model} is created by integrating a rosette trajectory for $10^{6}$~yrs, which corresponds to $\sim10$ orbital cycles of the CND and is closer to its estimated lifetime \citep{Sanders_1998, vollmer_2002, scoville_2005}. The phase coverage of this model is sparse, and thus its velocity field is sensitive to the position at which the orbit is launched. To account for this, we sample
different initial positions for the rosette by rotating it around the dynamical center through an angle $\phi_{\text{init}}$, which increases counter-clockwise from the $+x'{-}\text{axis}$ of the orbit's reference frame. The trajectory is always initiated at the pericenter. 

\textit{The Filled-Rosette model} is a rosette orbit with nearly complete phase coverage of the CND disk between $r_{\text{in}}$ and $r_{\text{out}}.$ It is constructed by evolving a single rosette trajectory for $10^{8}$~yrs. The longer integration period ensures that the final velocity profile is insensitive to where the orbit is initialized and there are no gaps in the model's on-sky area. The velocity field of this model is equally smooth as the Circular Disk model, but it shows a different spatial structure (see Section~\ref{sec:results_ros_m}). 

The set of parameters that are sampled for both rosette models are listed in Table~\ref{tab:params_sampled}. Orientation angles are discussed in Section \ref{sec:mock_obs}.

%%%%%%%%%%%%%%%%%%%%%%%%%%%%%%%%%%%%%%%%%%%%%%%%%%%%%%%%%%%%%%%%%%%%%%%
\subsection{On the NSC Potential}\label{sec:sphSymm_NSC}
%%%%%%%%%%%%%%%%%%%%%%%%%%%%%%%%%%%%%%%%%%%%%%%%%%%%%%%%%%%%%%%%%%%%%%%
The NSC is observed to be flattened along the Galactic plane (GP), with an axis ratio of $q \sim 0.80$ within $68''$ \citep{schodel_2014, Fritz_2016, Gallego-Cano_2020}. Additionally, at a projected radius of $1.7$\,pc ($42''$), the CND is estimated to be tilted by $5\degree$ relative to the GP. It increases up to a tilt of $25\degree$ for radii between $2.5$\,pc ($62''$) and $4.1$\,pc ($103''$) \citep{Genzel_1989, Genzel_2010}.

The bulk of the CND emission lies within approximately $-50''\leq \Delta\mathrm{R.A.} \leq 50''$ and $-70'' \leq \Delta\mathrm{decl.} \leq 70''$. This is the area to which we crop the observational map that is used to evaluate our models (see Section~\ref{sec:obsMap}). Thus, the data we fit is drawn from a region where the cluster exhibits only mild flattening and the tilt angle of the CND relative to the GP is small. Any effects due to cluster non-sphericity, such as nodal precession, are consequently not expected to have a significant impact on the gas velocities. We note that \cite{Sofue_2025} use the CND's kinematics to constrain the central potential and find that the disk's on-sky morphology and position-velocity diagrams are best reproduced by a spherical cluster. \cite{Gallego_Cano} and \cite{schodel_2018_II}, respectively, fit their stellar surface density and surface brightness maps of the NSC by assuming a spherical tracer density profile and report good fits to their data within 10\,pc (see Figure~14 of \cite{Gallego_Cano} and Figure~9 of \cite{schodel_2018_II}). Therefore, we presently adopt a spherical central potential and consider orbits that are aligned with the GP. The latter assumption is especially useful for both rosette models, since modeling the CND via multiple rosette segments on different planes introduces a large and highly degenerate parameter space. In future work, we intend to model the CND in an axisymmetric cluster and fit for its tilt relative to the GP.

%%%%%%%%%%%%%%%%%%%%%%%%%%%%%%%%%%%%%%%%%%%%%%%%%%%%%%%%%%%%%%%%%%%%%%%
\subsection{Motivating the Models}\label{sec:models_motivation}
%%%%%%%%%%%%%%%%%%%%%%%%%%%%%%%%%%%%%%%%%%%%%%%%%%%%%%%%%%%%%%%%%%%%%%%
The CND is expected to have formed from the tidal disruption of nearby giant molecular clouds (GMCs) and continues to be replenished by infalling gas \citep{Sanders_1998, Bradford_2005, Yusef_zadeh_2008b, Alig_2011, Oka_2011, Hsieh_2017, Hsieh_2021}. 

The Circular Disk model expands upon existing works in the literature that use one or a few circular orbits to describe the CND kinematics. It is an idealized representation of a gas disk in which the material is assumed to be uniformly distributed and fully circularized. Although emission lines from the CND show evidence of non-circular motion, its large-scale kinematics are dominated by circular rotation \citep{Guesten_1987, Bradford_2005, Tsuboi_2018}. This model thus provides a simple limiting-case description of the CND that is a useful baseline against which more complex configurations can be studied. 

Emission maps of the CND however show a clumpy disk within which numerous filaments are arranged in an intricate branching network \citep{scoville_2005, Martin_2012, Tsuboi_2018, Hsieh_2017}. Also, the CND's rotation is perturbed by large local velocity dispersions throughout the disk \citep{Guesten_1987, Hsieh_2021}. These properties are not reproduced by the smooth Circular Disk model. But, they are better captured by the Million-Years-Rosette model, which more closely resembles the observational map because its intertwined red- and blue-shifted rosette segments non-uniformly cover the CND's on-sky area and produce velocity substructure with steep gradients (see Section~\ref{sec:results_ros_m}).

The Filled-Rosette model represents a case where numerous rosette segments overlap to produce a smooth disk, with a velocity structure that is distinct from the one produced by nested circular orbits. Compared to the Circular Disk model, the Filled-Rosette shows higher velocity dispersions at each point in the disk, and its velocities are larger near the inner edge but are smaller otherwise, especially close to the outer edge (see Section~\ref{sec:results_ros_f}).

We note that neither rosette models account for the shocks that would be produced at points of self-intersection along the planar trajectory. Consequently, they do not describe the time evolution of the CND. Instead, they simply present a snapshot that approximates the disk’s present-day morphology and velocity field. For instance, the Million-Years-Rosette model can approximate the configuration in which a few mutually inclined, non-interacting streams superimpose in projection to form the CND, which can be motivated from studies that describe the CND using three kinematically distinct streamers (e.g, \cite{Jackson_1993, Martin_2012}). Ultimately, numerical simulations are necessary to properly model the temporal evolution and potential self-intersections of the streams.

%%%%%%%%%%%%%%%%%%%%%%%%%%%%%%%%%%%%%%%%%%%%%%%%%%%%%%%%%%%%%%%%%%%%%%%
\begin{deluxetable*}{rlllllllll}
%\tablewidth{25cm}
\tablecaption{List of the CND Model Parameters Sampled}
\label{tab:params_sampled}
\tablehead{
\colhead{Parameter} & \colhead{Circular Disk} & \colhead{Filled-Rosette} & \colhead{Million-Years-Rosette}
}
\startdata
Inner Radius (pc) & $0.5\leq r_{\text{in}} \leq2.0$ & $1.0\leq r_{\text{in}} \leq1.5$ & $1.0\leq r_{\text{in}} \leq1.5$ \\
Outer Radius (pc) & $2.0\leq r_{\text{out}} \leq4.0$ & $2.0\leq r_{\text{out}} \leq3.5$ & $2.0\leq r_{\text{out}} \leq3.5$ \\
Inclination (deg) & $40\leq i \leq80$ & $60\leq i \leq80$ & $50\leq i \leq80$ \\
Position Angle (deg) & $20\leq \mathrm{PA} \leq40$ & --- & --- \\
Longitude of Ascending Node (deg) & --- & $20\leq \Omega \leq40$ & $20\leq \Omega \leq40$ \\
Argument of Pericenter (deg) & 0.0 & $0\leq \omega < 360$ & {---} \\
Initialization Angle (deg) & --- & --- & $0\leq \phi_{\text{init}} < 360$ \\
Total Grid Points & 157,448 & 69,696 & 101,376 \\
\enddata
%%%%%%%%%%%%%%%%%%%%%%%%%%%%%%%%%%%%%%%%%%%%%%%%%%%%%%%%%%%%%%%%%%%%%%%
\tablecomments{We construct an $N$-dimensional parameter grid for each model. The Circular Disk model has $N=4$ parameters, two of which define the size of the disk in the orbital plane ($r_\mathrm{in}, $ $r_\mathrm{out}$), and the remaining two are angles ($i$, PA) that define its orientation in the plane of the sky. The Filled-Rosette and the Million-Years-Rosette models have $N = 5$ parameters, two of which define the size of the disk (the orbit's pericenter $r_\mathrm{in}$ and apocenter $r_\mathrm{out}$), while three specify the orientation angles ($i$, $\Omega,$ $\omega$) and ($i$, $\Omega,$ $\phi_{\mathrm{init}}$), respectively. For convenience, we have introduced $\phi_{\mathrm{init}}$, which is the angle towards the initial pericenter where the rosette orbit is initialized. It is connected to $\omega$ by the relation $\omega = 120\degree - \phi_{\mathrm{init}}.$ Rotation angles are discussed in Appendix~\ref{sec:app_A}. The grid spacing along each parameter axis is as follows: $\Delta r_\mathrm{in}=\Delta r_\mathrm{out}=0.1\,$pc, $\Delta i= \Delta \mathrm{PA} =1\degree,$ (Circular Disk model) and  $\Delta i= \Delta \Omega=2\degree$ (rosette models), and $\Delta \omega =\Delta \phi_{\mathrm{init}}=60\degree$. The total number of parameter sets that are sampled for each model type is listed in the bottom row. In total, we analyze $\sim3.3\times10^{5}$ model configurations.}
\end{deluxetable*}

%%%%%%%%%%%%%%%%%%%%%%%%%%%%%%%%%%%%%%%%%%%%%%%%%%%%%%%%%%%%%%%%%%%%%%%
\section{Comparison with observations}\label{sec:obs}
%%%%%%%%%%%%%%%%%%%%%%%%%%%%%%%%%%%%%%%%%%%%%%%%%%%%%%%%%%%%%%%%%%%%%%%

%%%%%%%%%%%%%%%%%%%%%%%%%%%%%%%%%%%%%%%%%%%%%%%%%%%%%%%%%%%%%%%%%%%%%%%
\subsection{Observational Map}\label{sec:obsMap}
%%%%%%%%%%%%%%%%%%%%%%%%%%%%%%%%%%%%%%%%%%%%%%%%%%%%%%%%%%%%%%%%%%%%%%%
We evaluate the quality of our models by comparing them with a velocity (moment-1) map of the CND in the HCN ($J=1{-}0$) line. This map is shown in Figure~\ref{fig:obs_panel_fig} and is one of the most detailed maps of the CND. We construct it using data from the Atacama Large Millimeter/submillimeter Array (ALMA) project 2017.1.00040.S. The data and the images from this project are publicly available through the ALMA archive \citep{Hsieh_2021}. The synthesized beam for this data is $3.''69\times \,2.''34$, with a position angle of $85\degree.2$. The cube has an RMS of 0.11~Jy/beam, and a spectral resolution of $0.8~\mathrm{km\,s^{-1}}$. While the data extends beyond the region plotted in Figure~\ref{fig:obs_panel_fig}, we crop it at $-50 ''\leq \Delta\mathrm{R.A.} \leq 50''$ and $-70'' \leq \Delta\mathrm{decl.} \leq 70''$ from the location of Sgr~A* to retain only the most prominent features of the CND. For comparison with our models, we only use the observational data with a signal-to-noise ratio $\sigma \geq 20.$

%%%%%%%%%%%%%%%%%%%%%%%%%%%%%%%%%%%%%%%%%%%%%%%%%%%%%%%%%%%%%%%%%%%%%%%
\subsection{Mock Observations}\label{sec:mock_obs}
%%%%%%%%%%%%%%%%%%%%%%%%%%%%%%%%%%%%%%%%%%%%%%%%%%%%%%%%%%%%%%%%%%%%%%%

The comparison of the models with the observations is conducted in the following way. After constructing the model in the CND plane, we rotate it into the plane of the sky using the matrix described in Appendix~\ref{sec:app_A}. The rotation angles are the inclination $i$, the longitude of ascending nodes $\Omega,$ and the argument of pericenter $\omega.$ For ease of comparison with other models in the literature, in the case of the Circular Disk model, we use the position angle (PA) to specify its orientation on the plane of the sky, instead of $\Omega$. PA is the angle between the $+y$-axis on the sky plane (i.e. north or $+$decl.) and the projected semi-major axis of the disk, measured east of north. Additionally, the argument of pericenter is set to $\omega = 0\degree$ for this model, since it is composed of concentric rings. 

For each model type, we sample $\sim 10^5$ sets of inner/outer radii and rotation angles. The parameter space of models we explore is given in Table~\ref{tab:params_sampled}.

After transforming the model onto the plane of the sky, we bin the model data onto a uniform position-velocity grid, whose sampling exactly matches that of the observational data. This  yields a data cube, with axes of R.A., decl., and $v_{z}$ (line-of-sight velocity). Each three-dimensional bin in this cube stores a count of the number of model points assigned to it. Next, we apply beam smearing. We convolve every velocity slice of the model cube with a two-dimensional elliptical Gaussian kernel whose FWHM matches the telescope beam. This produces a mock-observational data cube, from which we can construct our integrated emission map (moment-0), velocity map (moment-1), and velocity dispersion map (moment-2).

%%%%%%%%%%%%%%%%%%%%%%%%%%%%%%%%%%%%%%%%%%%%%%%%%%%%%%%%%%%%%%%%%%%%%%%
\subsection{Quantifying Model Accuracy}\label{sec:quantify_model}
%%%%%%%%%%%%%%%%%%%%%%%%%%%%%%%%%%%%%%%%%%%%%%%%%%%%%%%%%%%%%%%%%%%%%%%
Here, we describe the metric we use to quantify how well our models fit the observations. 

It might seem that the natural approach is to subtract the model velocity map from the observed one and calculate the root-mean-square (RMS) of the residual values. This however leads to a prominent issue: the CND velocity map is complex enough such that it is easy to find a wide variety of spurious models which fit a very tiny subset of the observational data extremely well but are completely unable to capture the overall structure. 

To avoid this issue, we devise the following set of requirements for the metric we use to quantify model accuracy. First, it needs to reward models with better spatial coverage of the observational map and punish those with a lack thereof. Second, it must assign a higher weight to models that better fit the prominent features of the observed map, which requires a weighting with respect to the total emission (moment-0) of the observational data. Third, as the observational map (Figure~\ref{fig:obs_panel_fig}) is somewhat patchy, we need to reward models that correctly capture the velocity features of the CND but might be slightly spatially offset from their observed locations.  

As the combination of the above requirements, we introduce the fit metric $(\mathcal{W})$ and the penalty factor $(\mathcal{P})$. Our best-fit model is the one that minimizes the product of these two quantities: 
\begin{equation}\label{eq:bestFit_criteria}
    {(\mathcal{W} \cdot \mathcal{P}) \to \min\, .}
\end{equation}

The fit metric $(\mathcal{W})$ is defined as
\begin{eqnarray}\label{eq:fit_metric}
\mathcal{W} = \frac{\mathfrak{D}}{\mathcal{N}}\, ,
\end{eqnarray}
where
\begin{subequations}\label{eq:RMS_metrics}
\begin{align}
    \label{eq:DE_RMS}
    \mathfrak{D} &= \frac{\sum_i \beta_{i}\, \cdot \,I_i}{\sum_i I_i}, \qquad \beta_{i} = \sqrt{\frac{\sum_j \Delta v^{2}_{ij}\, \cdot \,w_{ij}}{\sum_j w_{ij}}},
    \\
    \label{eq:vel_diffs}
    \Delta v_{ij} &= v_{\textrm{obs},\,i}\, - \,v_{\textrm{model},\,j} \; , \, \, 
    w_{ij} = \exp{-\frac{d^{2}_{ij}}{2\,\sigma_w^{2}}},
    \\
    d_{ij} &= \sqrt{(\mathrm{R.A.}_{i}\,-\,\mathrm{R.A.}_{j})^{2} + (\mathrm{decl.}_{i}\,-\,\mathrm{decl.}_{j})^{2}}\, ,
\end{align}
\end{subequations}%
and the normalization 
\begin{eqnarray}\label{eq:norm_factor}
\mathcal{N} = \left. \mathfrak{D} \right|_{v_{\mathrm{model},j}\,=\,0} =\frac{\sum_i \sqrt{v^{2}_{\textrm{obs},\,i}}\, \cdot \,I_i}{\sum_i I_i}\,,
\end{eqnarray}
with ``$v_{\mathrm{model},j}\,=\,0$'' defining a model with perfect coverage of the observational data but with all velocities set to zero, which is equivalent to having no model at all. 

Here, indices $i$ and $j$ iterate over points in the observational and model velocity maps, respectively; $v_{\mathrm{obs},i}$ are data points from the observed velocity map; $v_{\mathrm{model},j}$ are points from the model's mock-observational velocity map; $I_{i}$ is the integrated intensity (moment-0) corresponding to the $i^{\mathrm{th}}$ point in the observed velocity map; $d_{ij}$ is the distance between the points $i$ and $j$, measured in arcsec; and we set $\sigma_w = 2''$, which is equal to about half the resolution of the observations. When calculating the weights $w_{ij}$ for $\beta_i$, we only use points that lie within a radius of $d_{ij}\leq 25''$, as contributions from points with $d_{ij}> 25''$ to $\beta_i$ are negligible.

In this metric, the weights $w_{ij}$ allow us to evaluate whether a model correctly reproduces velocity trends in the observations. The weights $I_i$ reward models that provide a better fit to parts of the CND with stronger detected emission, as these areas yield more reliable constraints. The normalization $\mathcal{N}$ ensures that $\mathcal{W}$ is dimensionless.

The penalty factor $\mathcal{P}$ is defined as 
\begin{equation}
\label{eq:penalty_factor}
    \mathcal{P} = \sqrt{ \frac{1}{2}\left( \frac{N'}{N} \right)^{-2} + \frac{1}{2}\left( \frac{M'}{M} \right)^{-2} },
\end{equation}
where $N'$ is the number of points in the observational velocity map that have at least one model point within a radius of $r_{N}=0.''82$, $N$ is the total number of points in the observational velocity map, $M'$ is the number of points in the mock-observed model velocity map that have at least one observational point within a radius of $r_M=5''$, and $M$ is the total number of points in the mock-observed model velocity map. 

The first term under the square root in Equation~\ref{eq:penalty_factor} rewards models for covering as much data as possible, while the second term punishes the models for extending too far outside the observational map. The penalty factor is $\mathcal{P}\geq 1,$ with the minimal value of $\mathcal{P}=1$ for perfect coverage. In the cases where either $N'=0$ or $M'=0$, the penalty is infinite and the model is rejected. We use $r_{N}=0.''82$ to force models to fit the observational data closely. The chosen value of $r_N$ is equal to two grid spacings of the observational and mock-observed velocity maps. In contrast, we use a more generous value of $r_M=5''$ to avoid penalizing models for covering the gaps in the observational map (Figure~\ref{fig:obs_panel_fig}).
The value of $r_M$ is equal to about twice the observational beam size.

%%%%%%%%%%%%%%%%%%%%%%%%%%%%%%%%%%%%%%%%%%%%%%%%%%%%%%%%%%%%%%%%%%%%%%%
\subsection{Issues with an Unweighted Fit Metric}\label{sec:UWRMS_issues}
%%%%%%%%%%%%%%%%%%%%%%%%%%%%%%%%%%%%%%%%%%%%%%%%%%%%%%%%%%%%%%%%%%%%%%%
A metric similar to $\mathcal{W}$ in Equation~\ref{eq:fit_metric} but without any weighting factors--an unweighted velocity RMS--and with no penalty scheme might seem to be a more natural choice, particularly since the models and observations are on the same grid, which allows for easy subtraction of the two velocity maps. However, this approach has several major issues. 

First, an unweighted velocity RMS is strongly biased toward thin-ring models that fail to capture a majority of the CND's gas distribution, as it is generally easier to find a model that only fits a small subset of the data exceptionally well. We therefore have to make our metric prioritize better coverage of the CND's on-sky area. The role of the penalty factor is to strongly penalize models with poor spatial coverage.

Second, the observational map is somewhat patchy and is expected to have some contribution from both background and foreground emission. So, rather than fitting the smallest features in the data, we need to prioritize correctly fitting general trends at scales that encapsulate numerous observational points. Thus, we must make our metric prioritize models with better fits to general trends, even in the cases where the model values are slightly offset spatially from the observational points.

For example, while $\mathcal{W}$ is small for thin-ring models, their lack of spatial coverage leads to a high penalty factor $\mathcal{P}$. Combined, this leads to large ($\mathcal{W} \cdot \mathcal{P}$) values. So, as a measure of the model quality, we use the product $(\mathcal{W \cdot P})$, since it ties together how well a model captures both the velocities and spatial distribution of the CND. 

Last, the unweighted velocity RMS assigns the same weight to all residuals, as it does not know about which parts of the observational velocity map are detected more robustly. Instead, we want to make sure that models with lower residuals in regions of strong emission are prioritized. As such, we introduce weighting with respect to the integrated emission $I_i.$

%%%%%%%%%%%%%%%%%%%%%%%%%%%%%%%%%%%%%%%%%%%%%%%%%%%%%%%%%%%%%%%%%%%%%%%
\begin{deluxetable*}{rllllllllllc}
\tablecaption{Best-Fit Models and Their Parameters}
\label{tab:bestFit_params}
\tablehead{
\colhead{Model} & \colhead{$r_{\text{in}}$} & \colhead{$r_{\text{out}}$} & \colhead{PA} & \colhead{$i$} & \colhead{$\Omega$} & \colhead{$\omega$} & \colhead{$\phi_{\text{init}}$} & \colhead{$\bar{\delta v}$} & \colhead{$\mathcal{W}$} & \colhead{$\mathcal{P}$} &
\colhead{$(\mathcal{W}-\Delta\mathcal{W}_{SW})$}
\\ 
\colhead{} & \colhead{(pc)} & \colhead{(pc)} & \colhead{(deg)} & \colhead{(deg)} & \colhead{(deg)} & \colhead{(deg)} & \colhead{(deg)} & \colhead{(km\,s$^{-1}$)} & \colhead{} & \colhead{} & \colhead{}
}
\startdata
Circular Disk & 1.0 & 2.9 & 35 & 60 & --- & --- & --- & 12.4 & 0.37 & 1.61 & $\simeq 0.15$
\\
Filled-Rosette & 1.1 & 3.0 & 28.5$^{*}$ & 64 & 30 & 60 & --- & 20.4 & 0.41 & 1.61 & $\simeq 0.21$
\\
Million-Years-Rosette & 1.2 & 2.7 & 27.1$^{*}$ & 60 & 28 & --- & 180 & 22.2 & 0.41 & 1.66 & $\simeq 0.23$
\\
\enddata
\tablecomments{The first column lists the names of the different model classes (Section \ref{sec:models}). The rest of the columns list the corresponding best-fit parameters.
$r_{\text{in}}$/$r_{\text{out}}$ are the deprojected inner/outer radii defining the width of the CND;
disk orientation angles:
PA is the position angle,
$i$ is the inclination,
$\Omega$ is the longitude of ascending node,
$\omega$ is the argument of pericenter ($\omega = 0\degree$ for the Circular Disk model), and 
$\phi_{\mathrm{init}}$ is the angle in the orbital plane at which the rosette trajectory is initialized; 
$\bar{\delta v}$ is the emission-weighted median of the velocity residuals between the observational and model maps (Equation~\ref{eq:med_velOffset});
$\mathcal{W}$ is the fit metric (Equation~\ref{eq:fit_metric});
$\mathcal{P}$ is the penalty factor (Equation~\ref{eq:penalty_factor}), with $\mathcal{P} = 1$ implying perfect coverage;
and $(\mathcal{W}-\Delta\mathcal{W}_{SW})$ is the fit metric with residuals from the southwest region subtracted (Section \ref{sec:highResid_disc}).
The rotation matrix used to transform from the orbital plane to the plane of the sky is discussed in Appendix~\ref{sec:app_A}.
$^*$ denotes effective PAs calculated for the projected semi-major axes of the ellipses traced by the inner and outer edges of the rosettes.}
\end{deluxetable*}
%%%%%%%%%%%%%%%%%%%%%%%%%%%%%%%%%%%%%%%%%%%%%%%%%%%%%%%%%%%%%%%%%%%%%%%

%%%%%%%%%%%%%%%%%%%%%%%%%%%%%%%%%%%%%%%%%%%%%%%%%%%%%%%%%%%%%%%%%%%%%%%
\subsection{Velocity Offsets}
\begin{figure*}[ht!]
\centering
% height=0.85\textheight, keepaspectratio; width=0.95\textwidth
\includegraphics[width=0.99\textwidth, height=0.65\textheight, keepaspectratio]{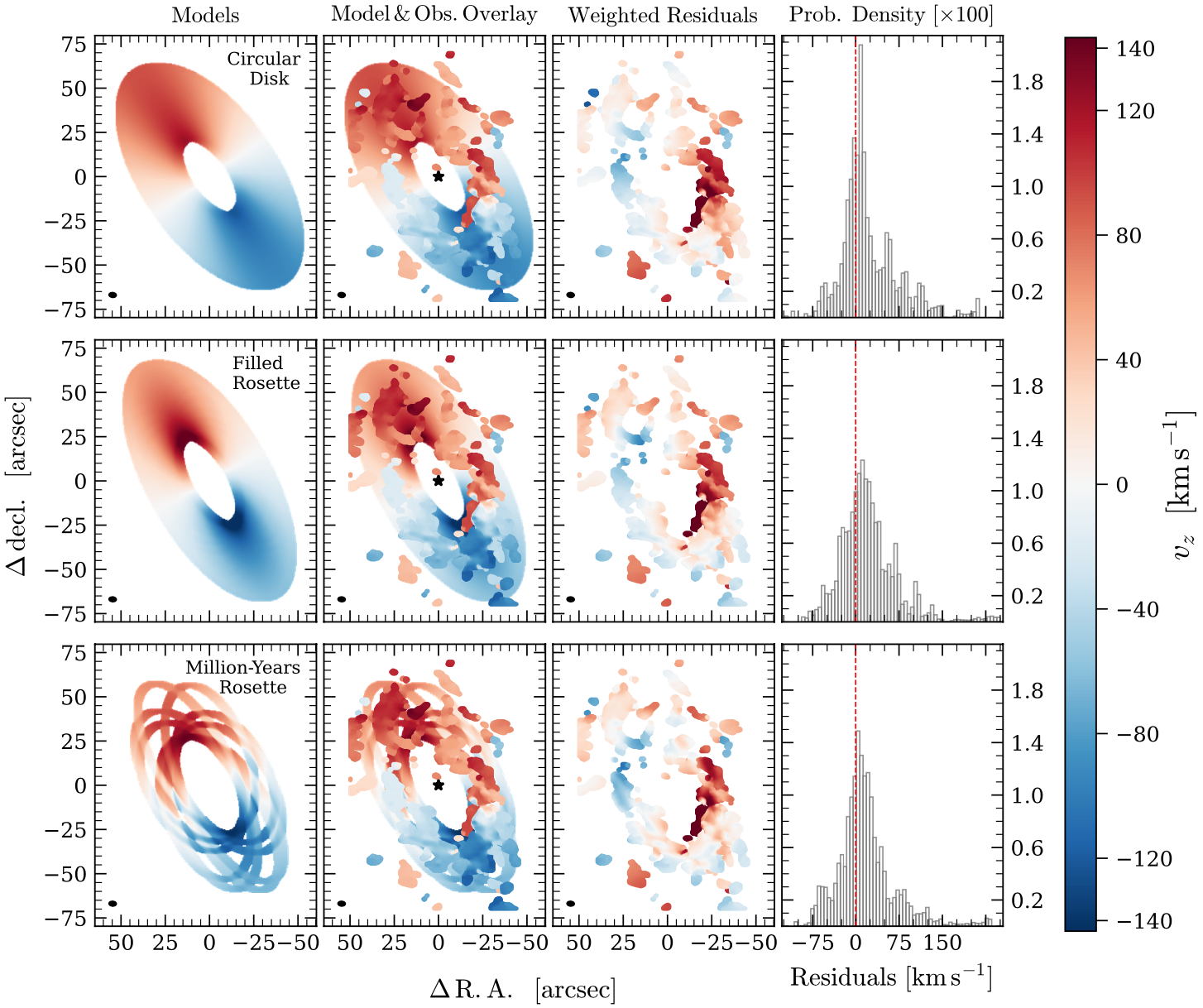}
\caption{Column 1 displays the mean velocity (moment-1) maps from each of our best-fit models. Column 2 overlays them with the observational velocity map (Figure~\ref{fig:obs_panel_fig}). Column 3 plots the distance-weighted average of the velocity residuals between the observational and model maps (see Equation~\ref{eq:panelFig_residuals}). Column 4 depicts these same residuals using normalized histograms, with their probability values multiplied by a factor of 100. The $\mathrm{x}{-}$axes and $\mathrm{y}{-}$axes in the first three columns indicate the $\mathrm{R.A.}$ and $\mathrm{decl.}$ offset from Sgr~A*, respectively. The bottom-left corner of these panels shows the kernel used to convolve the model data cubes, which is identical to the synthesized beam of the observational data \citep{Hsieh_2021}. The $\mathrm{x}{-}$axes in column 4 are velocity residuals in units of $\mathrm{km\,s^{-1}}$, and the red dotted lines mark $0\,\mathrm{km\,s^{-1}}$.} 
\label{fig:model_panel_fig}
\end{figure*}

%%%%%%%%%%%%%%%%%%%%%%%%%%%%%%%%%%%%%%%%%%%%%%%%%%%%%%%%%%%%%%%%%%%%%%%
We introduce a measure of the velocity offsets between the models and observations $(\bar{\delta v})$ that is unaffected by the large residuals from the southwest region of the CND, which are common across all models (see column 3 of Figure~\ref{fig:model_panel_fig}).

We use the weighted median of the velocity offsets
\begin{equation}\label{eq:med_velOffset}
\bar{\delta v} = \beta_{m}\, ,
\end{equation}
where $m$ is the index corresponding to the weighted median\footnote{Weighted median is defined as the $m^{\textrm{th}}$ element of the set $\{\beta_{i}\,I_i\}_\mathrm{sorted}$, which is $\{\beta_{i}\,I_i\}$ sorted in ascending order with respect to the $\beta_i$'s, with $m$ satisfying the following property: 
$m = \min_{k} \left[ \sum_{i=1}^{k} \{\beta_{i}\,I_{i}\}_{\textrm{sorted}} \; > \; \frac{1}{2} \sum_{i=1}^{N} \beta_{i}\,I_{i}\right].$} 
of the set $\{\beta_{i} \cdot I_i\}$, $\beta_i$ is defined in Equation~\ref{eq:DE_RMS}, and $I_i$ is the velocity-integrated intensity of the $i^{\textrm{th}}$ point. 

Previously, we defined $\mathfrak{D}$ as the weighted mean of $\beta_i$ with respect to the weights $I_i$ (Equation~\ref{eq:DE_RMS}). $\bar{\delta v}$ is simply its weighted median. The smaller the value of $\bar{\delta v}$, the smaller the median velocity offset between the model and the observational maps.

For the best-fit maps, for visualization purposes (see column 3 of Figure~\ref{fig:model_panel_fig}), we also calculate the sign-dependent analog of $\beta_i$: 
\begin{equation}\label{eq:panelFig_residuals}
    \mathcal{V}_{i} = \frac{\sum_j \Delta v_{ij} \cdot w_{ij}}{\sum_j w_{ij}}\, ,
\end{equation}
where $i$ corresponds to any single point on the observational map and $w_{ij}$ and $\Delta v_{ij}$ are defined in Equation~\ref{eq:vel_diffs}. As before, when calculating the sum and $w_{ij}$, we only use model points that are within a radius of $d_{ij}\leq 25''$ from each observational point $i$. Note that $\mathcal{V}$ is defined similarly to $\beta_i$ but without squaring $\Delta v_{ij}$ and without the overall square root.

%%%%%%%%%%%%%%%%%%%%%%%%%%%%%%%%%%%%%%%%%%%%%%%%%%%%%%%%%%%%%%%%%%%%%%%
%%%%%%%%%%%%%%%%%%%%%%%%%%%%%%%%%%%%%%%%%%%%%%%%%%%%%%%%%%%%%%%%%%%%%%%

%%%%%%%%%%%%%%%%%%%%%%%%%%%%%%%%%%%%%%%%%%%%%%%%%%%%%%%%%%%%%%%%%%%%%%%
\section{Results}\label{sec:discussion}
%%%%%%%%%%%%%%%%%%%%%%%%%%%%%%%%%%%%%%%%%%%%%%%%%%%%%%%%%%%%%%%%%%%%%%%
The best-fit models and their parameters for each of the three classes are presented in Figure~\ref{fig:model_panel_fig} and Table~\ref{tab:bestFit_params}. Below, we discuss these models in detail.

%%%%%%%%%%%%%%%%%%%%%%%%%%%%%%%%%%%%%%%%%%%%%%%%%%%%%%%%%%%%%%%%%%%%%%%
\subsection{Circular Disk Model}\label{sec:results_kep}
%%%%%%%%%%%%%%%%%%%%%%%%%%%%%%%%%%%%%%%%%%%%%%%%%%%%%%%%%%%%%%%%%%%%%%%
The best-fit model has $\mathcal{W} = 27.3\,\mathrm{km\,s^{-1}}/ \,73.9\,\mathrm{km\,s^{-1}} = 0.37$ and a penalty factor of $\mathcal{P}=1.61$. The first row of Figure~\ref{fig:model_panel_fig} shows that a circular disk accurately recovers the spatial profile of the CND and the broad blue-to-redshift gradient of its velocity field. The weighted velocity residuals $\mathcal{V}$ (Equation~\ref{eq:panelFig_residuals}) are plotted in column 3 of Figure~\ref{fig:model_panel_fig}. The probability density histogram of these residuals is depicted in column 4 of the same figure. It admits a narrow peak that is centered at $8.8\,\mathrm{km\,s^{-1}}$. This indicates that the Circular Disk model predominantly reproduces the general velocity pattern across the CND and has no systematic offset relative to the observations. 

However, the velocity field of this model is very smooth and has velocity dispersions (moment-2) in the range of $0{-}7.1\,\mathrm{km\,s^{-1}},$ which are considerably less compared to the $5{-}40\,\mathrm{km\,s^{-1}}$ dispersion values seen in observations \citep{Hsieh_2021,Oka_2011, Ferriere_2012, SW_2014, Genzel_2010}. The significant red (positive) residuals from the southwest region of disk, which are common across all three models, are discussed in Section~\ref{sec:highResid_disc}. 

%%%%%%%%%%%%%%%%%%%%%%%%%%%%%%%%%%%%%%%%%%%%%%%%%%%%%%%%%%%%%%%%%%%%%%%
\subsection{The Filled-Rosette Model}\label{sec:results_ros_f}
%%%%%%%%%%%%%%%%%%%%%%%%%%%%%%%%%%%%%%%%%%%%%%%%%%%%%%%%%%%%%%%%%%%%%%%
The best-fit Filled-Rosette model yields $\mathcal{W} = 0.41$ and a penalty factor of $\mathcal{P}=1.61$. The middle row of Figure~\ref{fig:model_panel_fig} shows that this model also accurately recovers the spatial profile of the CND and the broad blue-to-redshift gradient of its velocity field. The histogram of the weighted velocity residuals $\mathcal{V}$ (Equation~\ref{eq:panelFig_residuals}) peaks at $11.8\,\mathrm{km\,s^{-1}}$ and is fairly symmetric about this value. However, there are considerably fewer points with near $0\,\mathrm{km\,s^{-1}}$ velocity offsets compared to the Circular Disk model, and the histogram is broader. Section~\ref{sec:highResid_disc} discusses the high positive residuals from the southwest region, which produce the tail in the histogram. 

The Filled-Rosette model is similar to the Circular Disk model in terms of its spatial coverage, as indicated by the identical penalty factors. These models have almost the same inner and outer radii. However, the velocity structures of the model disks are distinctly different. The Filled-Rosette model has higher velocities at the inner edge of the disk, but typically lower velocities elsewhere. This is due to the averaging of overlapping red- and blue-shifted features from different segments of the trajectory (see the next section for more details). A comparison of these two models in Figure~\ref{fig:model_panel_fig} shows that while the Circular Disk overpredicts the velocities in some sections, the Filled-Rosette model underpredicts them. 

The Filled-Rosette model admits velocity dispersion (moment-2) values in the range $0{-}63\,\mathrm{km\,s^{-1}},$ which are closer to the observed dispersion value of $5{-}40\,\mathrm{km\,s^{-1}}$ \citep{Hsieh_2021,Oka_2011, Ferriere_2012, SW_2014, Genzel_2010}.

%%%%%%%%%%%%%%%%%%%%%%%%%%%%%%%%%%%%%%%%%%%%%%%%%%%%%%%%%%%%%%%%%%%%%%%
\subsection{The Million-Years-Rosette Model}\label{sec:results_ros_m}
%%%%%%%%%%%%%%%%%%%%%%%%%%%%%%%%%%%%%%%%%%%%%%%%%%%%%%%%%%%%%%%%%%%%%%%
The best-fit Million-Years-Rosette model results in $\mathcal{W} = 0.41$ and a penalty factor of $\mathcal{P}=1.66$. The bottom row of Figure~\ref{fig:model_panel_fig} confirms that this model correctly captures the spatial profile of the CND and reproduces the broad blue-to-redshift gradient as well as several substructures seen in its velocity field. The histogram of the weighted velocity residuals $\mathcal{V}$ (Equation~\ref{eq:panelFig_residuals}) appears to be approximately symmetric about its peak value of $4.3\,\mathrm{km\,s^{-1}}$. The positive-residual tail due to values in the southwest region is discussed in Section~\ref{sec:highResid_disc}. The histogram peak for this model is sharper than the Filled-Rosette model but less prominent than the Circular Disk model. Furthermore, the FWHM of the velocity residuals has a large tail compared to the Circular Disk, indicating that this model is less accurate at predicting the velocities.

The velocity map of this model is composed of rosette segments from $\sim 10$ orbital cycles, which leads to relatively sparse phase coverage compared to the Filled-Rosette model. As such, the substructure due to the averaging of overlapping segments is preserved, resulting in a complex velocity profile. The Million-Years-Rosette model has a velocity dispersion (moment-2) map that ranges between $0{-}53\,\mathrm{km\,s^{-1}},$ which is close to the observed CND dispersion of $5{-}40\,\mathrm{km\,s^{-1}}$ \citep{Hsieh_2021,Oka_2011, Ferriere_2012, SW_2014, Genzel_2010}.

%%%%%%%%%%%%%%%%%%%%%%%%%%%%%%%%%%%%%%%%%%%%%%%%%%%%%%%%%%%%%%%%%%%%%%%
\subsection{Residuals in the Southwest Region} \label{sec:highResid_disc}
%%%%%%%%%%%%%%%%%%%%%%%%%%%%%%%%%%%%%%%%%%%%%%%%%%%%%%%%%%%%%%%%%%%%%%%
Column 3 of Figure~\ref{fig:model_panel_fig} shows that the velocity residuals for all models peak near $-40''<\Delta\mathrm{R.A.}<-10''$ and $-30'' <\Delta\mathrm{decl.} < 20''$. In their corresponding histograms, this produces the secondary peaks close to $\pm75\,\mathrm{km\,s^{-1}}$ and the tail beyond $150\,\mathrm{km\,s^{-1}}$. 

This southwest region corresponds to the southern end of ``Anomaly A'' from \cite{Tsuboi_2018}. The authors identify this feature as a possible feeding channel for the CND and find that it indicates sites of strong shocks (velocity widths of $\sim50\,\mathrm{km\,s^{-1}}$). Hence, the gas dynamics in this region are likely affected by non-gravitational processes, which we do not consider. 

It is possible to remove this feature from the observational map prior to comparison with the models. However, its exact boundaries are not well defined. For consistency, we chose to retain it, since it punishes all models equally.

The approximate contribution of the southwest region to the fit metric $\mathcal{W}$ is about half. It is $\Delta \mathcal{W}_{SW} \simeq 0.22$ for the Circular Disk model, $\Delta \mathcal{W}_{SW} \simeq 0.20$ for the Filled-Rosette model, and $\Delta \mathcal{W}_{SW} \simeq 0.18$ for the Million-Years-Rosette model. The value of the fit metric with the southwest residuals removed is about $(\mathcal{W}-\Delta\mathcal{W}_{SW})\simeq 0.15$ for the Circular Disk model, $(\mathcal{W}-\Delta\mathcal{W}_{SW})\simeq 0.21$ for the Filled-Rosette model, and $(\mathcal{W}-\Delta\mathcal{W}_{SW})\simeq 0.23$ for the Million-Years-Rosette model.

%%%%%%%%%%%%%%%%%%%%%%%%%%%%%%%%%%%%%%%%%%%%%%%%%%%%%%%%%%%%%%%%%%%%%%%
\section{Conclusion}\label{sec:conclusion}
%%%%%%%%%%%%%%%%%%%%%%%%%%%%%%%%%%%%%%%%%%%%%%%%%%%%%%%%%%%%%%%%%%%%%%%
In this work, we conduct modeling of the dominant molecular gas structure in the Milky Way's Galactic Center---the CND---in the combined potential of the black hole and the nuclear star cluster. We consider three model classes: circular disks and non-Keplerian rosettes models (the Million-Years-Rosette and Filled-Rosette models). Our models are described in Section \ref{sec:models} and the sampled parameters are listed in Table \ref{tab:params_sampled}. With $\sim3.3\times10^{5}$ model configurations sampled, this is the most extensive modeling study of the CND to date.

To test model accuracy, we conduct mock observations for each of our models and compare them with an observational velocity map of the CND in the HCN ($J=1{-}0$) line, constructed from the data of \cite{Hsieh_2021}. Our mock observations account for 
the grid on which the observational data was imaged and for the effects of beam smearing. Then, we calculate the fit metric $\mathcal{W}$ (Equation~\ref{eq:fit_metric}), which measures how well the model reproduces the velocities seen in the observations, and the penalty factor $\mathcal{P}$ (Equation~\ref{eq:penalty_factor}), which determines how much of the CND's on-sky area the model captures. The best models are selected based on a combination of these measures (Equation~\ref{eq:bestFit_criteria}).

We find that all three classes of models can provide a good fit to the observational data (see Figure~\ref{fig:model_panel_fig} and Table~\ref{tab:bestFit_params}). The best model for the observational velocity map (moment-1) is achieved by a Circular Disk model with inner and outer radii of $r_{\textrm{in}} = 1.0\,$pc and $r_{\textrm{out}} = 2.9\,$pc, an inclination of $i = 60\degree$, and a position angle of $\textrm{PA} = 35\degree$. It outperforms both rosette models by about 50\% according to the fit metric without the residuals from the southwest region, which are common across all models (see $(\mathcal{W}-\Delta\mathcal{W}_{SW})$ values in Table \ref{tab:bestFit_params}). For the velocity dispersion properties of the CND, the closest fit is achieved by the Million-Years-Rosette model with  $r_{\textrm{in}} = 1.2\,$pc and $r_{\textrm{out}} = 2.7\,$pc, an inclination of $i = 60\degree$, and an effective position angle of $\textrm{PA} = 27\degree.1$ (Section~\ref{sec:discussion}).

This work significantly expands the set of CND models available in the literature, which until now primarily consisted of single-ring models. Since the Circular Disk model is favored, our results suggest that a majority of the gas in the CND may be nearly circularized. We note that all models considered here are simplified representations of the observed CND. Numerical simulations are required to account for the temporal evolution and possible collisions of the CND streams, which will be the subject of our future works. 

%%%%%%%%%%%%%%%%%%%%%%%%%%%%%%%%%%%%%%%%%%%%%%%%%%%%%%%%%%%%%%%%%%%%%%%

%%%%%%%%%%%%%%%%%%%%%%%%%%%%%%%%%%%%%%%%%%%%%%%%%%%%%%%%%%%%%%%%%%%%%%%
\section*{Acknowledgements}
%%%%%%%%%%%%%%%%%%%%%%%%%%%%%%%%%%%%%%%%%%%%%%%%%%%%%%%%%%%%%%%%%%%%%%%

We thank the anonymous referee for their suggestions, which helped to improve this manuscript. A.U. would like to thank Mark Morris, Yoram Lithwick, and Dani Lipman for helpful discussions and their insightful comments. 

M.D.G. was partially supported by the CIERA Postdoctoral Fellowship from the Center for Interdisciplinary Exploration and Research in Astrophysics at Northwestern University.

This work used computing resources provided by Northwestern University and the Center for Interdisciplinary Exploration and Research in Astrophysics (CIERA). This research was supported in part through the computational resources and staff contributions provided for the Quest high performance computing facility at Northwestern University, which is jointly supported by the Office of the Provost, the Office for Research, and Northwestern University Information Technology.

This paper makes use of the following ALMA data: ADS/JAO ALMA 2017.1.00040.S. ALMA is a partnership of ESO (representing its member states), NSF (USA), and NINS (Japan), together with NRC (Canada), MOST and ASIAA (Taiwan), and KASI (Republic of Korea), in cooperation with the Republic of Chile. The Joint ALMA Observatory is operated by ESO, AUI/ NRAO, and NAOJ.

%%%%%%%%%%%%%%%%%%%%%%%%%%%%%%%%%%%%%%%%%%%%%%%%%%%%%%%%%%%%%%%%%%%%%%%

%% Bibliography
%\bibliography{references}{}
%\bibliographystyle{aasjournalv7-2}



\bibliography{references}{}
\bibliographystyle{aasjournalv7-2}

%%%%%%%%%%%%%%%%%%%%%%%%%%%%%%%%%%%%%%%%%%%%%%%%%%%%%%%%%%%%%%%%%%%%%%%

%%%%%%%%%%%%%%%%%%%%%%%%%%%%%%%%%%%%%%%%%%%%%%%%%%%%%%%%%%%%%%%%%%%%%%%
\appendix
\section{Rotation to the plane of the sky}\label{sec:app_A}
%%%%%%%%%%%%%%%%%%%%%%%%%%%%%%%%%%%%%%%%%%%%%%%%%%%%%%%%%%%%%%%%%%%%%%%
The orbits for all models are initialized in a reference frame that is defined using a right-handed coordinate system $x'y'z'$, with the dynamical center located at the origin. The orbital plane is the $x'y'$ plane. Trajectories in the orbital plane rotate about the dynamical center in a counterclockwise direction. Since the potential is spherical, all orbits remain in the $x'y'$ plane. 

On the plane of the sky, the $x$-axis is positive along east, indicating the direction of increasing R.A. and the $y$-axis is positive along north, indicating the direction of increasing decl. The positive $z$-axis is then oriented away from the observer. 

In both coordinate systems, the axes measure the distance relative to Sgr~A*. To rotate the coordinates and velocities from the orbital plane to the plane of the sky, the following matrix is used: 
\begin{equation}
\label{eq:rotMat_orbSky}
\mathrm{T} =
\begin{bmatrix}
\quad \sin\Omega \,\cos\omega + \cos\Omega \,\cos i \,\sin\omega & &
-\sin\Omega \,\sin\omega + \cos\Omega \,\cos i \,\cos\omega & &
\cos\Omega \,\sin i &
\\
\quad \cos\Omega \,\cos\omega - \sin\Omega \,\cos i \,\sin\omega & &
-\cos\Omega \,\sin\omega - \sin\Omega \,\cos i \,\cos\omega & &
\sin\Omega \,\sin i &
\\
\quad \sin i \,\sin\omega & &
\sin i \,\cos\omega & &
-\cos i &
\end{bmatrix}, 
\end{equation}
where the inclination $i$ is measured from the $-z$-axis to the angular momentum vector of the orbit; $\Omega$ is measured from the $+y$-axis (north/$+$decl.) and increases counter-clockwise, as viewed from the $-z$ direction, toward the vector that connects the dynamical center to the ascending node, which is where the orbit crosses the sky plane from the negative to positive $z$-axis; and the argument of pericenter $\omega$ is measured from the ascending node to the pericenter of the orbit, increasing along the direction of motion. For the rosette models, because the orbits exhibit apsidal precession, $\omega$ does not correspond to the classical Keplerian element. Instead, it simply encodes the location of the initial periapsis about the force center, from which the orbit is launched.   

For the Circular Disk model, we set $\omega = 0\degree$ and, instead of $\Omega$, use the position angle (PA) of its projected semi-major axis to define the disk's orientation in the sky. The PA is measured east of north. For the Million-Years-Rosette model, for simplicity, we use $\phi_{\textrm{init}}$ instead of $\omega$. $\phi_{\textrm{init}}$ is the angle toward the first pericenter at which the rosette orbit is initialized. These angles are connected by the relation $\omega = 120\degree - \phi_{\textrm{init}}$.

%%%%%%%%%%%%%%%%%%%%%%%%%%%%%%%%%%%%%%%%%%%%%%%%%%%%%%%%%%%%%%%%%%%%%%%%

\end{document}